\documentclass[aps,prl,superscriptaddress,twocolumn,10pt,nofootinbib,preprintnumbers]{revtex4}

\usepackage{graphicx}
\usepackage{epstopdf}
\usepackage{bm}
\usepackage{epsfig}
\usepackage{graphics}
\usepackage{amsmath}
\usepackage{xcolor}

\begin{document}

\title{Electrocrystallization of Supercooled Water in Confinement}

\author{\firstname{R~M.}~\surname{Khusnutdinoff}}
\email{khrm@mail.ru}
\affiliation{Kazan Federal University, 420008 Kazan, Russia}
\affiliation{Udmurt Federal Research Center of the Ural Branch of the Russian Academy of Sciences, 426068, Izhevsk, Russia}

\author{\firstname{A.~V.}~\surname{Mokshin}}
\email{anatolii.mokshin@mail.ru}
\affiliation{Kazan Federal University, 420008 Kazan, Russia}
\affiliation{Udmurt Federal Research Center of the Ural Branch of the Russian Academy of Sciences, 426068, Izhevsk, Russia}


\begin{abstract}
The paper discusses the features of supercooled water thin film of width $d=3.97$~nm contained by the perfect graphene layers and crystallizing under external stationary electric field. It was found that the electric field applied perpendicular to graphene layers impedes structural ordering, while the electric field applied in lateral direction contributes to formation of the cubic ice ($Ic$) phase, which is thermodynamically less stable compared to the hexagonal ice ($Ih$) phase. It is shown that the growth of the $Ic$ crystalline phase occurs without formation of intermediate crystalline phases. It was found that the crystallization rate depends strongly on the magnitude of the applied electric field. In particular, the processes of full electrocrystallization of the system do not appear over simulation time scale ($\sim 40$~ns) if the electric field of the magnitude less than $0.07~\rm{V/\AA}$ is applied.
\end{abstract}
\maketitle

\section{Introduction}
Bulk water can form various phases, including liquid and vaporous phases, as well as numerous crystalline and amorphous phases of ice \cite{r1,r2}. When water is at specific spatial confinements, namely, adsorbed at interfaces or enclosed in microscopic pores \cite{r3}, then it can be characterized by the physical properties different from these for bulk states. Such low-dimensional water determines the aspects of various phenomena in materials science, nanotechnology, geology, and biology. The results of numerical simulations and theoretical calculations predict many possible thermodynamic phases for adsorbed and confined water, while experimental verification of these results is currently problematic. Recently, water layer was experimentally studied of nanometer size enclosed between two graphene layers and it was found that under these specific conditions, water forms the crystalline ice phase with the square lattice \cite{r4}. These results are very intriguing because the detected phase has symmetry qualitatively different from the usual tetrahedral geometry of hydrogen bonds typical for water. In fact, the ice phase with the square lattice can be considered as a separate layer of the cubic ice in the crystallographic plane (001). Note that this crystalline phase with the square lattice is characterized by a high packing density with the lattice constant equal to $2.83~\rm{\AA}$. This work develops the ideas proposed in \cite{r4} and is aimed to study crystallization of thin film of supercooled water.

\section{MODEL AND SIMULATION DETAILS}
The simulated system presents a water film of a nanometer size enclosed between two perfectly smooth graphene layers. The interaction between water molecules was carried out on the basis of the Tip4p/Ice model potential, which properly describes the phase diagram and properties of crystalline and amorphous ices \cite{r5}. The intramolecular bonds and angles are constrained by conditions according to the SHAKE algorithm. To take into account the long-range Coulomb interactions between the partial charges, we used the PPPM-method with the cutoff radius $r_c=13~\rm{\AA}$. The interaction between water molecules and carbon atoms was carried out using the Lennard-Jones potential, where the interaction parameters were determined on the basis of the Lorentz-Berthelot mixing rule \cite{r6}.

The simulation was performed in NVT ensemble for the supercooled water at the temperature $T=268$~K and density $\rho=0.92$~g/cm$^3$. To stabilize the temperature of the system, the Nose-Hoover thermostat with a relaxation parameter of $1.0$~ps was applied. Periodic boundary conditions were applied along the $X$ and $Y$ directions. Schematic sketch of the simulated system is presented in Fig. 1a.

\begin{figure}[h!]
\begin{center}
\includegraphics[width=\linewidth]{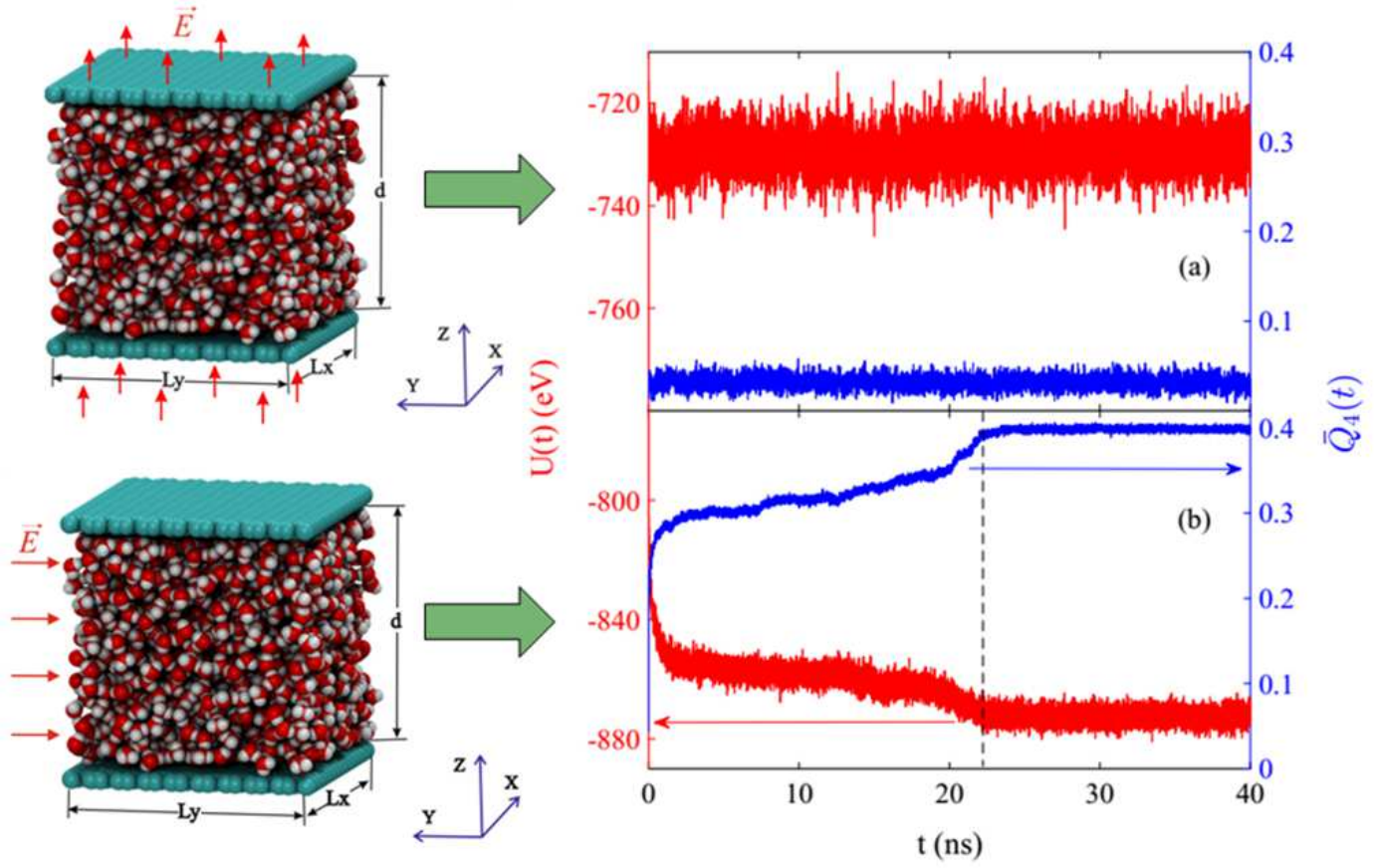}
\caption{(a) Schematic sketch of the simulated system; (b, c) time dependences of the internal energy and global orientational order $\bar{Q}_{4}$, for confined water film at the temperature $T=268$~K and density $\rho=0.92$~g/cm$^3$ under the external electric field of the magnitude $E=0.5~\rm{V/\AA}$.}
\end{center}
\end{figure}
We independently considered two cases when an external stationary electric field is applied perpendicular to graphene layers and parallel to graphene layers. The imposed electric field was stationary, and the situations with various magnitudes $E$ of this field, in a range of $0.05\div0.5~\rm{V/\AA}$, were simulated. The statistical averaging of the evaluated quantities was performed over results of ten independent simulation runs.

\section{RESULTS AND DISCUSSION}
To identify the effect of the field on the structural ordering, we use the structural analysis, which is based on calculation of the global Steinhardt-Nelson-Ronchetti orientational order parameter \cite{r7}:
\begin{equation}
\bar{Q}_{l}=\left(\frac{4\pi}{2l+1}\sum_{m=-l}^{l}\left|\frac{\sum\sum Y_{lm}(\theta_{ij},\varphi_{ij}) }{\sum N_{b}(i)}\right|^{2}\right)^{1/2}.
\label{eq_Q6_bop}
\end{equation}
Here, $N_b(i)$ is the number of nearest neighbors for the $i$th molecule, $Y_{lm}(\theta_{ij},\varphi_{ij})$ are the spherical harmonics; $\theta_{ij}$ and $\varphi_{ij}$ are polar and azimuthal angles, respectively. It should be noted that each type of crystal lattice is characterized by a unique set of values of the orientational order parameters $\bar{Q}_{l}$ (where $l=4, 6, 8, \ldots$). This allows to identify local crystalline structures of a certain type. So, for example, the parameter $\bar{Q}_{4}$ takes values $0.259$ and $0.506$ for perfect hexagonal and cubic ices, respectively. For disordered systems (water, amorphous ice), the parameters $\bar{Q}_{4}$ and $\bar{Q}_{6}$ take values close to zero.

For the case of the electric field imposed perpendicular to the simulated system, no structural ordering is observed. The internal energy $U$ and the order parameter $\bar{Q}_{4}$ do not change with time (Fig.~1b). On the other hand, as found, the field applied in lateral direction promotes structural ordering. Here, the process of structural ordering is determined by two characteristic regimes. The initial regime is associated with the orientation of dipole water molecules under the influence of an external electric field. The subsequent regime is characterized by relaxation of the metastable system into the crystalline phase. As seen in Fig.~1c, the time scale of electrocrystallization of the sample is evaluated as $\tau=23\pm 1$~ns.

\begin{figure}[h!]
\begin{center}
\includegraphics[width=\linewidth]{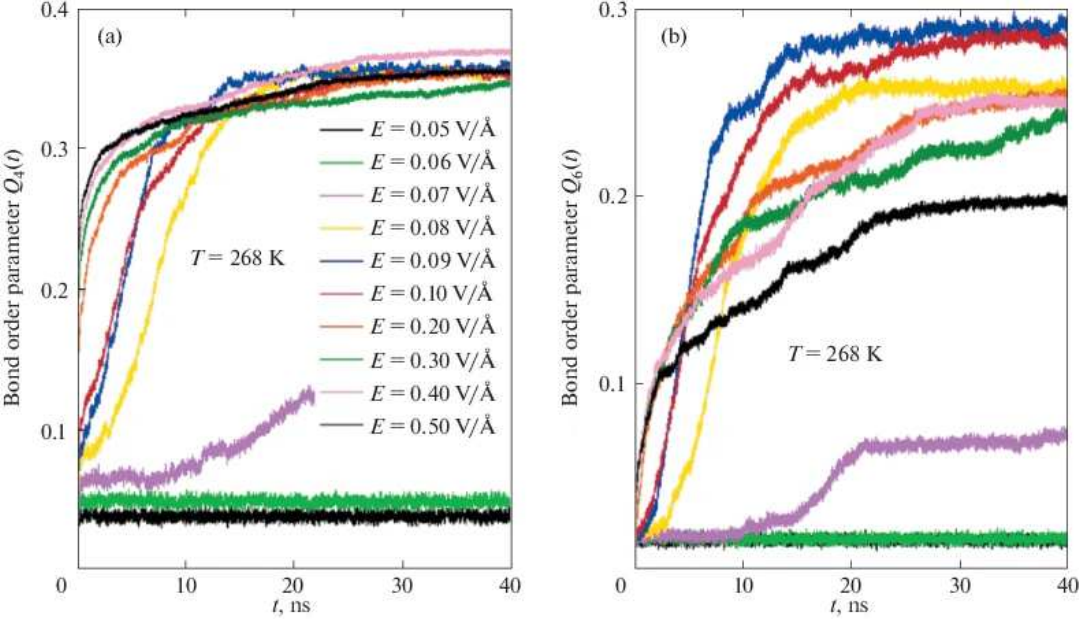}
\caption{Time dependences of the global orientational order parameter (a) $\bar{Q}_{4}$ and (b) $\bar{Q}_{6}$, for water film in confinement at the temperature $T=268$~K and density $\rho=0.92$~g/cm$^3$ for various magnitudes of the external electric field.}
\end{center}
\end{figure}
Now we consider how the magnitude of the electric field does impact on the structural ordering. For this, the time dependences of the order parameters were calculated for crystallization at various electric field strengths $E$ in a range of $0.05\div0.5~\rm{V/\AA}$. As seen from Fig.~2, for the case with the electric field magnitudes $E<0.07~\rm{V/\AA}$, the structural ordering is not observed over simulation time scales ($\sim 40$~ns). On the other hand, the fields of higher magnitudes, $E\geq 0.07~\rm{V/\AA}$, promote the crystallization.

\begin{figure}[h!]
\begin{center}
\includegraphics[width=\linewidth]{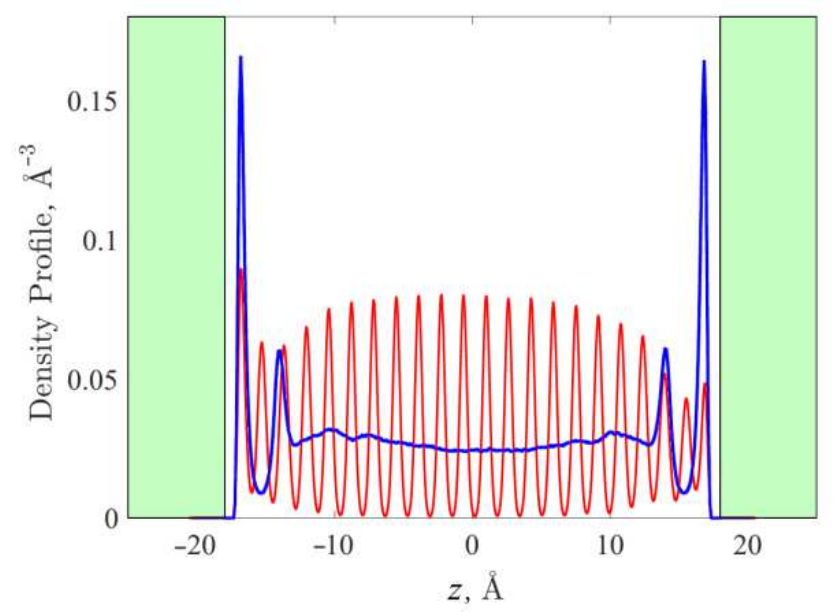}
\caption{Density profile of the system at the temperature $T=268$~K. Thick line corresponds to the system in the absence of a field; thin line presents results for the system under electric field of the magnitude $E=0.1~\rm{V/\AA}$ at the time moment $\tau=40$~ns after start of applying the field.}
\end{center}
\end{figure}
Emergent crystalline phase can be also detected by the specific contour of the density profile. In Fig.~3, the density profiles are shown, evaluated for the system under study in the absence of the field (thick line) and for the system under external field with the magnitude $E=0.1~\rm{V/\AA}$ (thin line).

In the absence of the field, a layering is observed only for the spatial ranges near the graphene walls, that is due to the hydrophobicity of the walls. On the other hand, for the case of imposed electric field, the periodic contour of the density profile, typical for an ordered structure, is observed.

\begin{figure}[h!]
\begin{center}
\includegraphics[width=\linewidth]{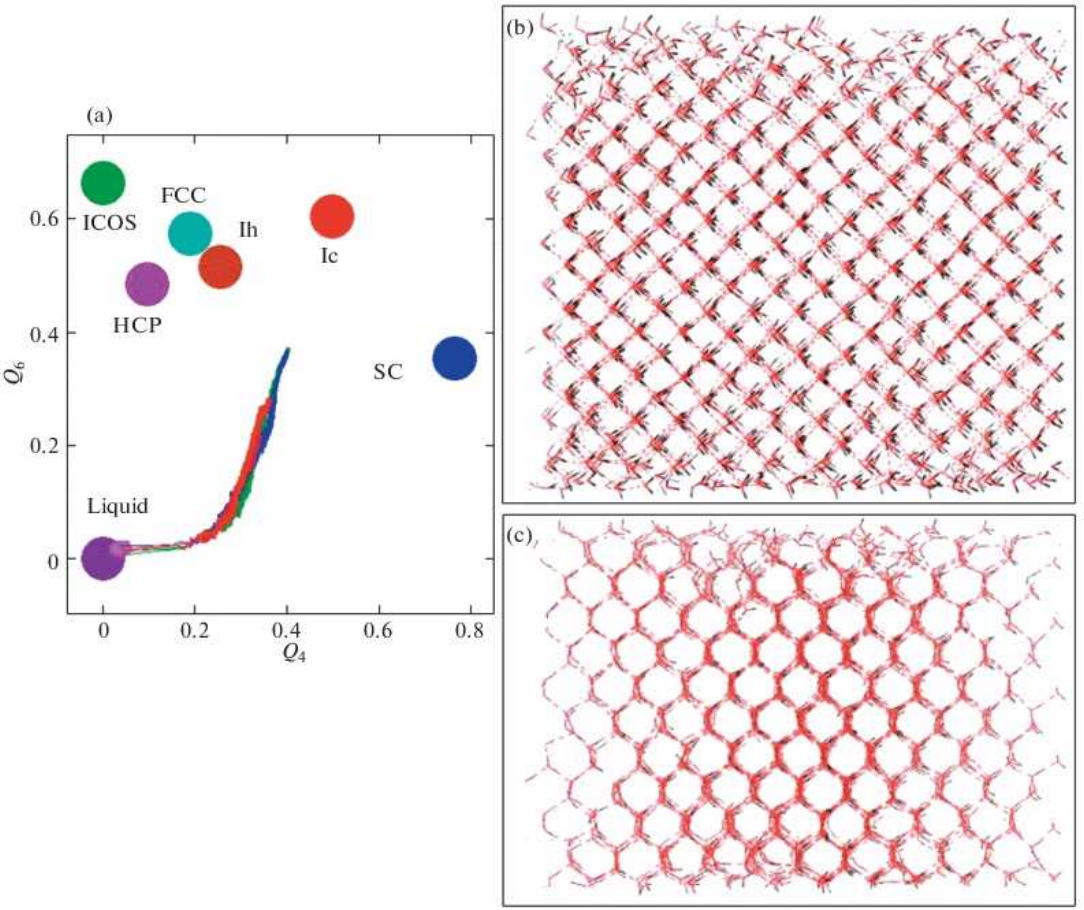}
\caption{(a) Trajectories of formation of the ordered phase on the ($\bar{Q}_{4}$, $\bar{Q}_{6}$) diagram for the system at the temperature $T=268$~K and the density $\rho=0.92$~g/cm$^3$ for various magnitudes of the external electric field. Colored circles indicate location of various possible crystalline phases on the diagram. (b, c) Snapshots of the system under study in the crystallographic planes (001) and (101), respectively.}
\end{center}
\end{figure}
In Fig.~4, we show ($\bar{Q}_{4}$, $\bar{Q}_{6}$) diagram with the trajectories that demonstrate formation of the ordered phase at various magnitudes of the electric field.

\section{SUMMARY}
In the present work, using molecular dynamics simulations, it was found that the external stationary electric field acting on thin film of supercooled water confined by graphene walls promotes crystallization. As found, the structural ordering of the system depends strongly on the direction and magnitude of the applied electric field. Namely, the field applied in the lateral direction leads to the formation of cubic ice phase that is less thermodynamically stable in comparison with hexagonal ice phase. It was established that formation of the $Ic$ phase occurs without formation of intermediate crystalline phases.

\section{Funding}
This work was supported by the Russian Foundation for Basic Research (project no. 18-02-00407-a). The molecular dynamic simulations were performed by using the computational cluster of Kazan Federal University and the computational facilities of Joint Supercomputer Center of the RAS.

\end{document}